\documentclass{article}
\usepackage{spconf,amsmath,graphicx}

\usepackage[usenames,dvipsnames]{xcolor}

\newcommand{\MSEt}[0]{\xi_T}
\newcommand{\X}[0]{\mathbf{X}}

\newcommand{\D}[0]{\mathbf{D}}
\newcommand{\Y}[0]{\mathbf{Y}}

\renewcommand{\P}[0]{\mathbf{P}}
\newcommand{\Pest}[0]{\hat{\mathbf{P}}}
\newcommand{\M}[0]{\mathbf{M}}
\newcommand{\N}[0]{\mathbf{N}}
\newcommand{\E}[0]{\mathbf{E}}

\newcommand{\Ce}[0]{\mathbf{C}_e}

\newcommand{\Exp}[1]{\mbox{E}\{ #1 \}}

\newcommand{\diag}[1]{\mbox{diag}\left\{ #1 \right\}}
\newcommand{\trace}[1]{\mbox{Tr}\left\{#1\right\}}

\newcommand{\fpool}[1]{d_{#1}}
\newcommand{\fpoolv}[0]{\mathbf{d}}
\renewcommand{\sharp}[0]{\boldsymbol{\nu}}
\newcommand{\sharpi}[1]{\nu_{#1}}
\newcommand{\avdelay}[0]{\bar{\delta}}
\newcommand{\srates}[0]{\boldsymbol{\mu}}
\newcommand{\srate}[1]{\mu(#1)}

\newcommand{\DFT}[1]{\lambda_{#1}}
\newcommand{\DFTpool}[0]{\mathbf{\Lambda}_{d}}

\newcommand{\fpoolopt}[0]{\mathbf{d}_{\text{opt}}}
\newcommand{\fpoollong}[0]{\mathbf{d}_{\text{long}}}
\newcommand{\fpoolshort}[0]{\mathbf{d}_{\text{short}}}

\newcommand{\consset}[0]{\mathcal{C}}

\newcommand{\Rxx}[0]{\mathbf{R}_{xx}}
\newcommand{\Rxyx}[0]{\mathbf{R}_{xyx}}

\newcommand*{\Fitcol}[1]{\resizebox{0.45\textwidth}{!}{$#1$}}

\title{Filter design for delay-based anonymous communications}
\name{Simon Oya$^{\star}$ \qquad Fernando P\'erez-Gonz\'alez$^{\star}$ \qquad Carmela Troncoso$^{\dagger}$\thanks{S.~Oya is funded by the Spanish Ministry of Education, Culture and Sport under the FPU grant. F.~P\'erez-Gonz\'alez is funded by the Spanish Ministry of Economy and Competitiveness and the ERDF under projects TACTICA, COMPASS (TEC2013-47020-C2-1-R) and COMONSENS (TEC2015-69648-REDC), and by the Galician Regional Government and ERDF under projects GRC2013/009 and AtlantTIC. C.~Troncoso is supported by EU H2020-ICT-10-2015 NEXTLEAP (GA n 688722)}}
\address{$^{\star}$Signal Theory and Communications Dept., University of Vigo\\
      $^{\dagger}$The IMDEA Software Institute}

\begin{document}
\graphicspath{{./img/}}

%
\maketitle
\begin{abstract}
 In this work, we address the problem of designing delay-based anonymous communication systems. We consider a timed mix where an eavesdropper wants to learn the communication pattern of the users, and study how the mix must delay the messages so as to increase the adversary's estimation error. We show the connection between this problem and a MIMO system where we want to design the coloring filter that worsens the adversary's estimation of the MIMO channel matrix. We obtain theoretical solutions for the optimal filter against short-term and long-term adversaries, evaluate them with experiments, and show how some properties of filters can be used in the implementation of timed mixes. This opens the door to the application of previously known filter design techniques to anonymous communication systems.
\end{abstract}
\begin{keywords}
Anonymity, filter design, timed mixes, optimization
\end{keywords}
\section{Introduction}
\label{sec:intro}

Anonymity in communication systems is typically achieved at the expense of delay or communication bandwidth. Mixes~\cite{Chaum81,DiazSerjantovMixes}, the basic building blocks of high-latency anonymous communication systems, are channels that \emph{delay} messages, change their appearance and output them in a random order in batches. This confuses an eavesdropper trying to unveil the path followed by the messages in the network, who is not able to identify the sender of a message leaving the mix with absolute certainty.

It is well known that two delaying mechanisms that cause the same average delay in the communication can achieve different protection against a malicious observer. Since delay is the main resource to generate privacy in delay-based anonymous communication systems, it is of paramount importance to understand how to use it optimally.

The design of the \emph{delya characteristic} of an anonymous communication system has been studied in the literature with different privacy goals in mind~\cite{DanContMixes,RebolloPool,TNET16}. In this work, we take over the work in~\cite{TNET16}, where the authors obtain the optimal delay characteristic against an eavesdropper with global vision of the network that tries to learn the average number of messages each sender sends to each receiver. We start from the results in that paper and interpret the problem as a filter design problem in a MIMO system, where the delay characteristic is a filter, which allows us to reason about some results obtained in~\cite{TNET16}. We also make a clear distinction between two scenarios that give different optimal delay characteristics, and find the optimal characteristic in an scenario where senders send most of their messages to only one of their friends, which is not covered by previous work. Finally, we show how filter properties can help in the implementation of delay functions in a decentralized way, which is of particular interest in practice.

The rest of the document is distributed as follows. Section~\ref{sec:preliminaries} presents the system model and notation, as well as previous results that are relevant for this work. It also sets up the optimization problem of designing the delay characteristic that maximizes the privacy of the users. We solve this problem in Sect.~\ref{sec:design} for different scenarios, one of which was not considered in previous works, and validate our results. Finally, in Sect.~\ref{sec:utility} we give examples of how some properties of filter design can be used in the implementation of delay-based anonymous communications, and conclude in Sect.~\ref{sec:conclusions}.

\section{Preliminaries}
 \label{sec:preliminaries} 
 
 \subsection{System Model and Notation}
 We consider a system where a group of $N$ senders, indexed by $i\in\{1,\cdots,N\}$, send messages to a group of $M$ receivers, indexed by $j\in\{1,\cdots,M\}$, through an anonymous communication channel, which we model as a \emph{timed mix}. The timed mix contains a timer that loops continuously, counting down starting at $\tau$ seconds, thus creating laps that are called \emph{rounds}. When a message arrives at the mix, it is assigned a random delay drawn from a probability mass function called \emph{delay characteristic}. When the timer expires, all the messages whose delay is zero have their appearance changed through cryptographic tools and are forwarded to their corresponding recipients. The messages that remain in the mix have their delay decreased in one unit, and will leave eventually as the timer loops.

\begin{figure}[t]
\begin{minipage}[b]{\linewidth}
  \centering
  \centerline{\includegraphics[width=8.5cm]{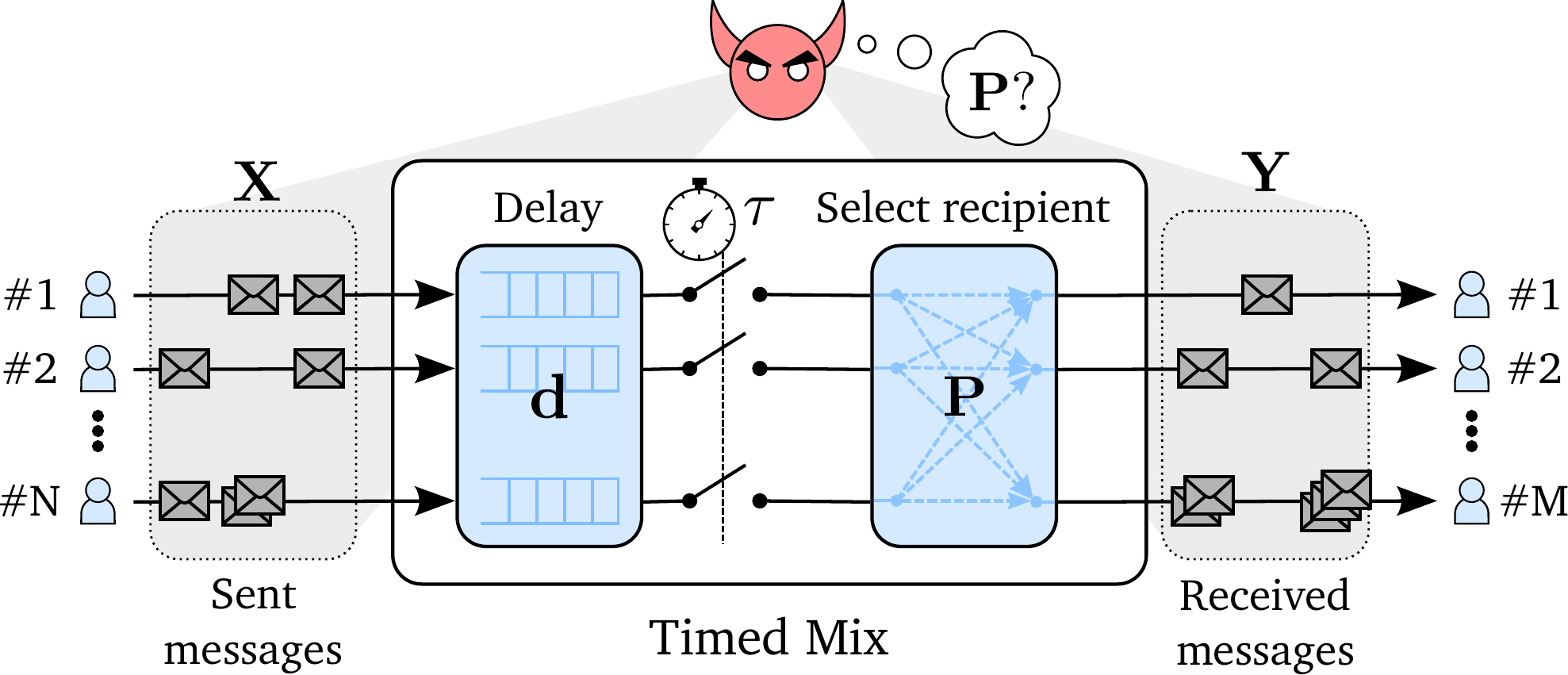}}
\end{minipage}
\caption{System model. The adversary observes $\X$ and $\Y$, knows $\fpoolv$ and $\tau$, and wants to learn the probabilities in $\P$.}
\label{fig:mix}
\end{figure}

 We consider an adversary observing the messages arriving and leaving the mix for $T$ seconds, i.e., during a total of $\rho=\lfloor T/\tau \rfloor$ rounds. The adversary can observe all the messages sent by every sender and received by every recipient (i.e., it is \emph{global}), and it is a mere observer of the system (i.e., it is \emph{passive}). One example of such adversary is a malicious Internet Service Provider. The goal of this adversary is to learn the probability that a message sent by $i$ is received by $j$, denoted by $p_{j,i}$. This represents the percentage of messages from $i$ that go to $j$ on average. We assume that the adversary knows how the mix works (i.e., $\tau$ and the delay characteristic) but cannot look inside it. The change of appearance of the messages inside the mix prevents the adversary from performing bit-wise linkability of input and output messages, while the random delay and grouping of messages in rounds prevents timing linkability. This system is depicted in Fig.~\ref{fig:mix}.
 
 We now introduce the statistical model and the notation, which is summarized in Table~\ref{tab:notation}. The random variable that models the number of messages sent by user $i$ in round $r$ is denoted by $X_i^r$.   
 Each input message is delayed independently and randomly according to the delay characteristic $\fpoolv\doteq[\fpool{0},\fpool{1},\cdots,\fpool{\rho-1}]^T$, where $\fpool{k}$ is the probability that the mix delays a message $k$ rounds. The recipient of each message sent by user $i$ is $j\in\{1,\cdots,M\}$ with probability $p_{j,i}$. Finally, the total number of messages that receiver $j$ gets in round $r$ is $Y_j^r$. Using these variables, we also build the $\rho\times N$ matrix of all input observations $\X$, whose $i,r$-th element is $X_i^r$, i.e., $(\X)_{r,i}=X_i^r$. Likewise, we build $\Y$ and $\P$ as $(\Y)_{r,j}=Y_j^r$ and $(\P)_{i,j}=p_{j,i}$.
 
 \begin{table}
\begin{center}
\caption{Summary of notation}
\label{tab:notation}
\begin{tabular}{r l | p{0.55cm} p{1cm}}
  $p_{j,i}$ & Prob.~user $i$ sends a message to $j$. &  $(\P)_{i,j}$&$=p_{j,i}$. \\
  $\fpool{k}$ & Prob.~a msg.~is delayed $k$ rounds. & $(\fpoolv)_{k}$&$=\fpool{k}$. \\
  $X_i^r$ & No.~msgs.~sent by $i$ in round $r$. & $(\X)_{r,i}$&$=X_i^r$. \\
  $Y_j^r$ & No.~msgs.~received by $j$ in round $r$. & $(\Y)_{r,j}$&$=Y_j^r$. \\
  $\sharpi{i}$ & ``Sharpness'' of $i$, $\sharpi{i}=\sum_j p_{j,i}^2$. & $(\sharp)_k$&$=\sharpi{i}$.
 \end{tabular}
\end{center}
\end{table}

 \subsection{Previous Results and Connection with MIMO Communications}
 Previous works have studied different problems in this same model. Here, we summarize some results, mainly from \cite{TNET16,TIFS14}, that are of particular interest for our work. The first result is that $\Exp{\Y|\X}=\D\cdot\X\cdot\P$, where $\D$ is a convolution matrix defined as $(\D)_{r,s}=\fpool{r-s}$ if $r-s\geq 0$, and 0 otherwise. This means that, on average, the $M$ output processes in the columns of $\Y$ are a linear combination (produced by $\P$) of the $N$ input processes that are the columns of $\X$, convolved with the delay characteristic $\fpoolv$. If we define the noise of the outputs $\N\doteq\Y-\D\X\P$, another result is that the covariance matrix of this noise is $\mathbf{\Sigma}_\mathbf{\N|\X}=\diag{\D\X\mathbf{1}_N}-\D\cdot\diag{\X\sharp}\cdot\D^T$. In this expression, $\mathbf{1}_N$ is an $N\times 1$ all-ones vector, and $\sharp$ is a vector whose $i$-th entry is $\sharpi{i}\doteq\sum_{j=1}^M p_{j,i}^2$. This parameter represents the ``sharpness'' of the sending behavior of user $i$. Values $\sharpi{i}\approx 1$ represent a sender that focuses in a single receiver, while $\sharpi{i}\approx 0$ represents a sender that distributes messages evenly among her recipients. 
 
 From these results, we can see the $\fpoolv$ box in Fig.~\ref{fig:mix} as a linear filter and $\P$ as a MIMO channel matrix for the averages, and consider that $\Y$ is obtained after adding the noise $\N$ to this average. The problem of designing the delay characteristic $\fpoolv$ against an adversary that wants to estimate the sending behavior $\P$ is then equivalent to the problem of designing a \emph{filter} $\fpoolv$ against an adversary that wants to estimate the MIMO channel matrix $\P$. We will use this interpretation of $\fpoolv$ as a filter below. Note that even though $\X$ and $\Y$ only contain integer numbers (because messages cannot be broken in smaller units), with this interpretation we can disregard the integer constraints because $\X$ and $\Y$ are observations, not parameters to estimate.
 
 \subsection{Privacy Metric and Analysis}
 As mentioned above, the adversary wants to learn the probabilities $\P$, that are sensitive information from the users, after observing $\X$ and $\Y$ and knowing $\tau$ and $\fpoolv$. Previous works delve into the study of the best linear estimator for $\Pest$ in this scenario and the privacy metric used to assess the success of this adversary. We refer to~\cite{TNET16,TIFS14} for a thorough derivation of these results, and just note that the overall mean squared error $\MSEt$ of the adversary can be written as $\MSEt=\trace{\M\Ce\M}$, where $\M\doteq\diag{[\srate{1},\cdots,\srate{N}]}$ is a normalizing diagonal matrix and $\Ce$ is the covariance matrix of the adversary's error $\Ce\doteq\Exp{\E\E^T}$, where $\E\doteq\Pest-\P$. This matrix can be written for the best linear estimator of $\P$ as
  \begin{equation} \label{eq:Ce}
   \Ce=\Exp{(\X^T\D^T\D\X)^{-1}\X^T\D^T\mathbf{\Sigma}_{\N|\X}\D\X(\X^T\D^T\D\X)^{-1}}\,.
  \end{equation}
 Using the expression for $\mathbf{\Sigma}_{\N|\X}$ above, we have a relation between the overall privacy $\MSEt$ and the delay characteristic $\fpoolv$.

 \section{Design of the Optimal Delay Characteristic}
 \label{sec:design}
 
 Now we study how to design the delay characteristic $\fpoolv$ such as to increase the adversary's overall error $\MSEt$. In this section, we use $\DFT{k}$ to denote the $k$-th DFT coefficient of the $\rho$-point DFT of $\fpoolv$. We start by discussing some constraints on $\fpoolv$ and set up the design problem that gives us the optimal filter $\fpoolopt$. Then, we study the solution to this problem in two cases: 1) when the adversary observes the system indefinitely ($\rho\to\infty$) and $\rho\gg N$, and 2) when the adversary observes the system a number of rounds $\rho$ commensurate to $N$. This distinction, which is not clear in previous works, is important because it leads to different solutions. We call each of these scenarios long-term and short-term adversary/attack, respectively. 
 
 \subsection{Constraints on the Delay Characteristic}
 \label{sec:constraints}
 
 Although it is possible to see the delay characteristic $\fpoolv$ as a filter, we must keep in mind that it is actually a probability mass function, and therefore it must follow some constraints:
 
  {\bf a) Non-negativity:} since the values of this filter are probabilities, it must hold that $\fpool{k}\geq0$ for all $k$. The consequences of non-negativity constraints in filter design have been discussed in \cite{liu2010new,liu2010frequency,liu2010fundamental}. An immediate consequence is that $\DFT{0}\geq\DFT{k}$ for all $k$. A more complex effect of these constraints, explained in detail in \cite{liu2010frequency}, is that it is easier to achieve a filter $\fpoolv$ with large attenuation factor in high frequencies than in middle and low frequencies.
  
  {\bf b) Normalization:} another direct consequence of the fact that $\fpoolv$ is a probability mass function is $\sum_{k=0}^{\rho-1} \fpool{k}=1$. This forces the first DFT coefficient to be one, i.e., $\DFT{0}=1$.
  
  {\bf c) Average delay:} we want to design a mix that guarantees that the average delay of the messages in the system, measured in rounds, does not exceed some value $\avdelay$, i.e., $\sum_{k=0}^{\rho-1}k\cdot\fpool{k}\leq \avdelay$. One of the effects of this constraint is that, of all the filters that give the same magnitude response, we will prefer the minimum phase solution, i.e., the one whose zeros lie inside the unit circle. This comes from the fact that the group delay of the filter and the average delay $\avdelay$ are closely related.
  
 We denote the space of filters $\fpoolv$ that follow these constraints by $\consset$. This is defined formally as
 \begin{equation}
  \consset=\left\{\fpoolv: \sum_{k=0}^{\rho-1}\fpool{k}=1, \sum_{k=0}^{\rho-1} k\cdot\fpool{k}\leq\avdelay, \fpool{k}\geq 0\quad\forall k \right\}\,.
 \end{equation}
 Note that the filters $\fpoolv\in\consset$ are causal (by definition) and stable. Then, the optimal delay characteristic we are looking for, denoted by $\fpoolopt$, is the solution to
  \begin{equation} \label{eq:optpool}
  \fpoolopt=\underset{\fpoolv\in\consset}{\text{argmax }}\trace{\M\Ce\M}\,.
\end{equation}
 where $\Ce$ is given in \eqref{eq:Ce}. We now study this solution against long-term and short-term adversaries.

 \subsection{Long-Term Optimal Delay Characteristic}
 
When $\rho\to\infty$ and  $\rho\gg N$, we can approximate the terms $\X^T\D^T\D\X$ and $\X^T\D^T\mathbf{\Sigma}_{\N|\X}\D\X$ in \eqref{eq:Ce} by their expected values and obtain a closed form expression for $\MSEt$. Let $\gamma_1\doteq \sum_k \fpool{k}^2$ and $\gamma_2\doteq \sum_r \left(\sum_k \fpool{r} \fpool{r+k} \right)^2$. If we assume the inputs are i.i.d.~Poissonian, in the Appendix we show that when users have several friends with whom they communicate evenly ($\sharpi{i}\approx 0$), then $\MSEt$ grows with $1/\gamma_1$. When users send most of their messages to only one of their friends ($\sharpi{i}\approx 1$), then $\MSEt$ grows with $(\gamma_1-\gamma_2)/\gamma_1^2$. Therefore, we set
 \begin{equation} \label{eq:poollong}
  \fpoollong=\begin{cases}
              \underset{\fpoolv\in\consset}{\text{argmax }} (1/\gamma_1) &\text{ if }\sharpi{i}\approx 0\,,\\
              \underset{\fpoolv\in\consset}{\text{argmax }} (\gamma_1-\gamma_2)/\gamma_1^2&\text{ if }\sharpi{i}\approx 1\,.
             \end{cases}
 \end{equation}
 The study of $\fpoollong$ when $\sharpi{i}\approx 0$, including a close-form expression for it, already appears in \cite{TNET16}. However, it is important to note that when $\sharpi{i}\approx 1$ the solution changes, and this case was not analyzed in~\cite{TNET16}. Note that we can also extend this solution to the case where the input samples are not independent by including a coloring filter $\mathbf{g}$ before the adversary observation in the model, and replacing $\fpoolv$ by $\fpoolv * \mathbf{g}$ in the formulas above.
 
 Now we evaluate these results, comparing the numerical solution $\fpoolopt$ using \eqref{eq:optpool} and the theoretical one $\fpoollong$ in \eqref{eq:poollong}. We generate $\X$ using real data fed to a timed mix. We use the real data in~\cite{TNET16} so that this work is comparable to previous ones (see~\cite{TNET16} for a thorough description of the real datasets). We take $N=100$ users and $\rho=1\,500$ rounds from the datasets in order to study a case where $\rho\gg N$, and generate $\P$ using a Zipf distribution with 10 friends per sender for $\sharpi{i}\approx 0$ and a single friend for $\sharpi{i}=1$, with $M=100$. We compute $\fpoolopt$ and $\fpoollong$ for the 3 datasets in~\cite{TNET16} and show the average results in Fig.~\ref{fig:experiments}.a. We can see that the optimal filter obtained analytically \eqref{eq:poollong} is very close to the numerical one obtained through evaluation of \eqref{eq:optpool} using $\X$.

 \subsection{Short-Term Optimal Delay Characteristic}
\begin{figure}[t]
\begin{minipage}[b]{\linewidth}
  \centering
  \centerline{\includegraphics[width=4.0cm]{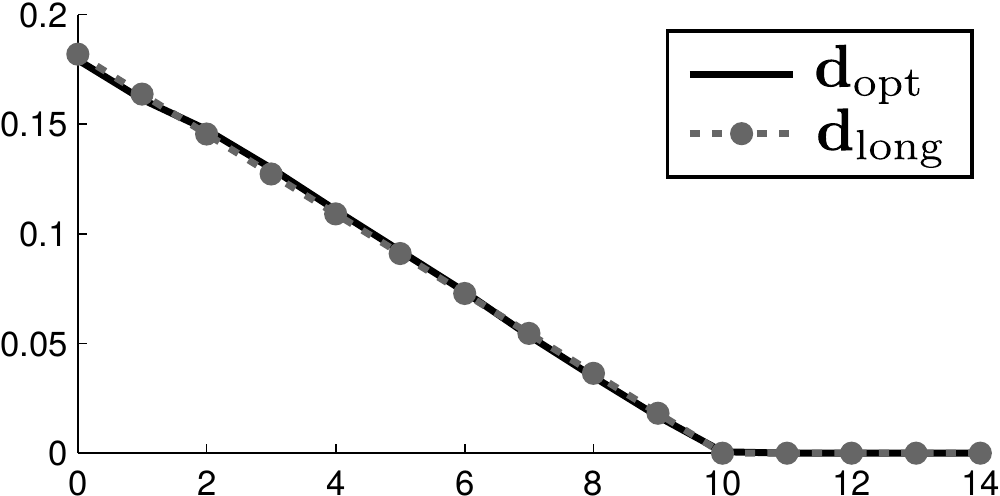}\quad \includegraphics[width=4.0cm]{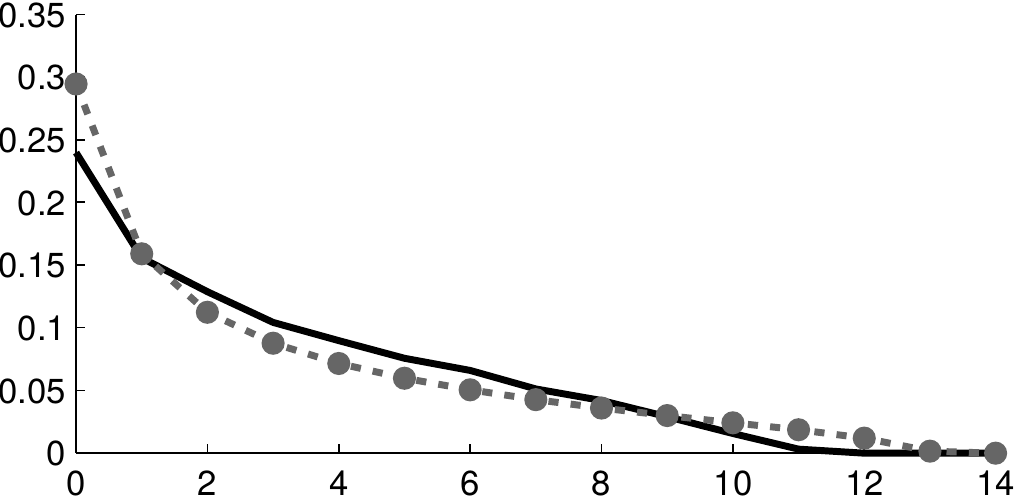}}
  \centerline{(a) $\fpoolv$, long-term adversary, $\sharpi{i}\approx 0$ (left) and $\sharpi{i}=1$ (right).}\medskip
\end{minipage}
\begin{minipage}[b]{\linewidth}
  \centering
  \centerline{\includegraphics[width=4.0cm]{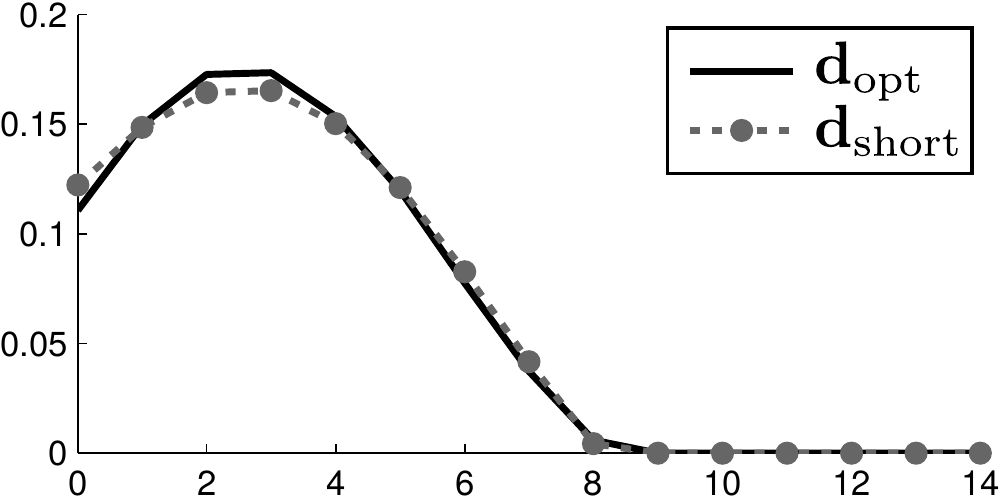}\quad \includegraphics[width=4.0cm]{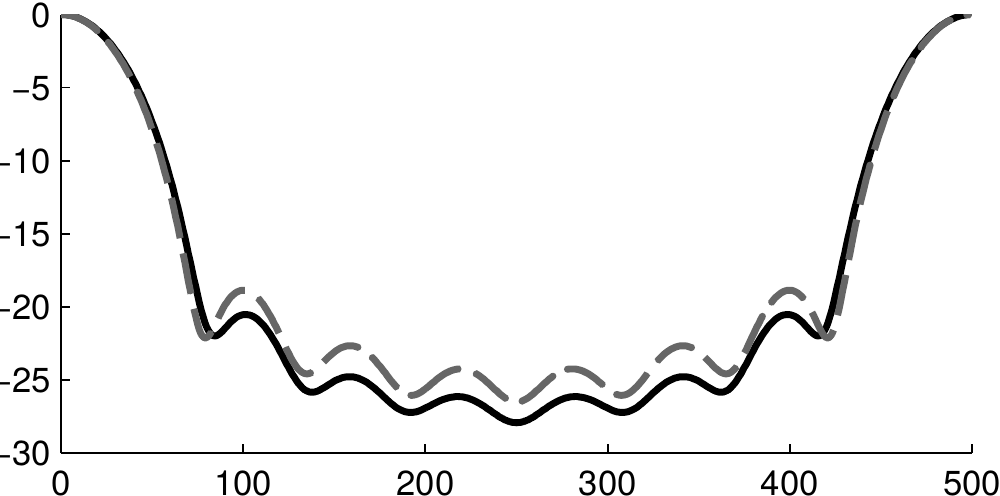}}
  \centerline{(b) $\fpoolv$ (left) and DFT of $\fpoolv$ (right), short-term adversary, $\sharpi{i}\approx 0$.}\medskip
\end{minipage}
\caption{Optimal delay characteristic against a long-term ($\rho=1\,500$) and short-term ($\rho=500$) adversary, with $N=100$.}
\label{fig:experiments}
\end{figure}

 We now consider the case where the number of observed rounds $\rho$ is commensurate with the number of senders $N$. In this case, it is argued in~\cite{techrep2016} that $\X^T\D^T\D\X$ should be made as close to singular as possible. We can write this as $\X_f^H|\DFTpool|^2\X_f$ where $\X_f$ is the $\rho\times N$ matrix of the DFT of the inputs and $\DFTpool$ is the diagonal matrix containing the coefficients of the $\rho$-point DFT of $\fpoolv$. To make this matrix close to singular, it makes sense to try to make $\rho-N$ DFT coefficients of $\fpoolv$ close to zero. This is easier to achieve for high-frequencies, as a consequence of the positivity constraints \cite{liu2010frequency}. Therefore, we can set\footnote{This is for even $N$. A small tweak is required for odd $N$.}
  \begin{equation} \label{eq:poolshort}
   \fpoolshort=\underset{\fpoolv\in\consset}{\text{argmin }} \sum_{k=N/2+1}^{\rho-N/2+1} \DFT{k}\,.
\end{equation}
The results obtained numerically (generating $\X$ by taking $N=100$ users and $\rho=500$ rounds from the real datasets) using \eqref{eq:poollong} and analytically with \eqref{eq:poolshort} are close, as shown in Fig.~\ref{fig:experiments}.b. From a filter-design perspective, the frequency response in dB (Fig.~\ref{fig:experiments}.b, right) confirms that the optimal pool is the low-pass filter that tries to remove information from the $\rho-N=400$ high-frequency DFT coefficients.

 \section{Applications}
 \label{sec:utility}
 
 In this section, we show how some properties of filters can be used to aid in the implementation of anonymous communication systems. Normally, anonymous communication systems are designed in a non-centralized way, with several devices connected in a network, for scalability and trust reasons~\cite{DDM03,MCPS03,gulcubabel,JAP}. In this case, the delay characteristic of the timed mix is the addition of the individual delays provided by the different components of the network.
 The task of designing a decentralized delay-based anonymous communication system can be simplified by considering basic properties of filter design. We illustrate the application of some of these properties (direct form implementation, cascade and parallel filters) in the two examples below and leave the study of how these findings can be used to configure and improve real systems like Mixmaster~\cite{MCPS03} and JAP~\cite{JAP} for future work.
 
\begin{figure}[t]
\begin{minipage}[b]{.48\linewidth}
  \centering
  \centerline{\includegraphics[width=4.0cm]{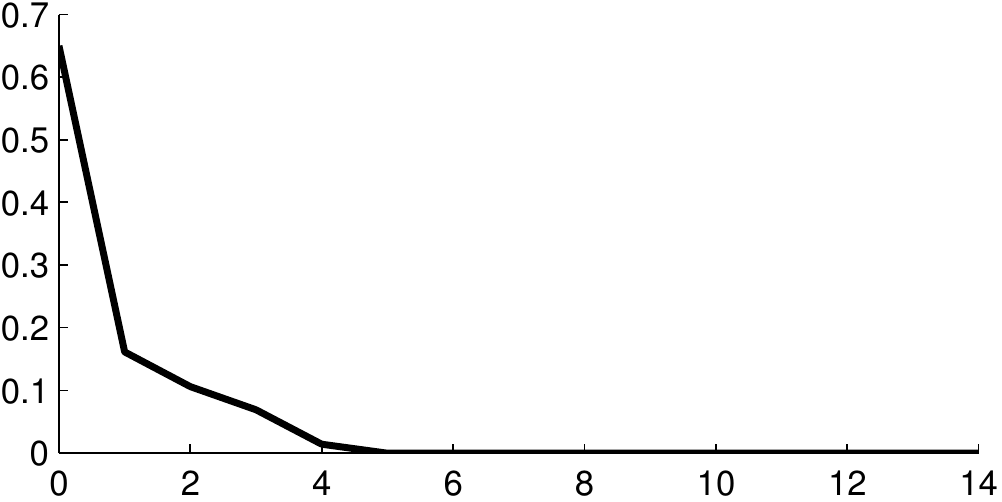}}
  \centerline{(a) $\fpoolv$, single node.}\medskip
  \centerline{\includegraphics[width=4.0cm]{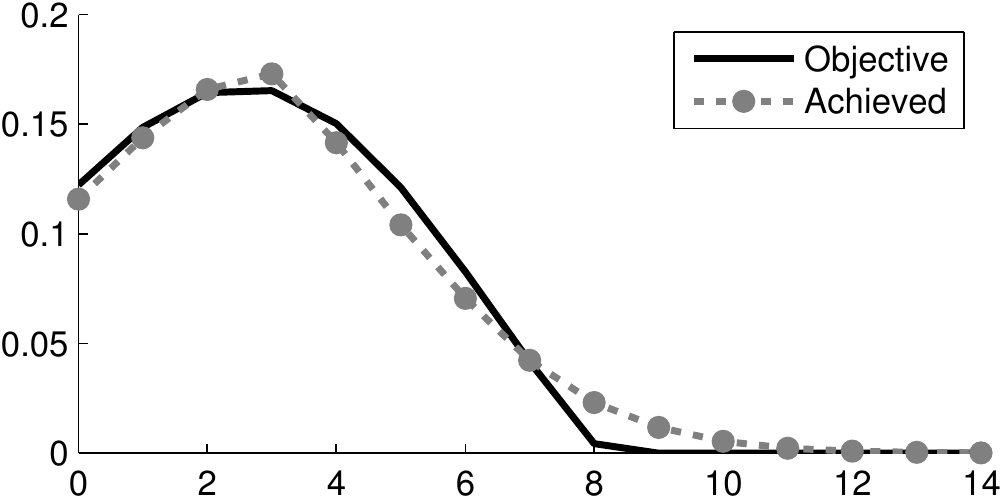}}
  \centerline{(b) $\fpoolv$ achieved vs.~objective.}\medskip
  \centerline{\includegraphics[width=4.0cm]{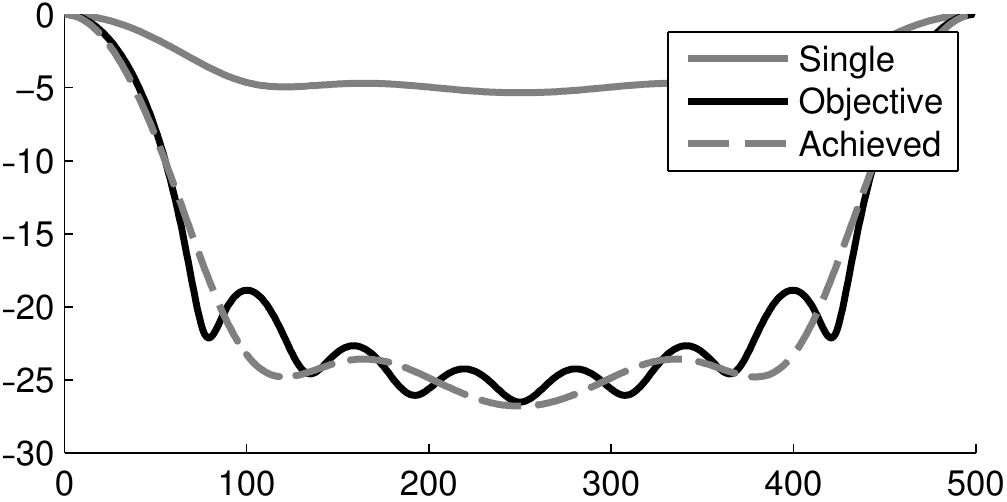}}
  \centerline{(c) Frequency response.}\medskip
\end{minipage}
\begin{minipage}[b]{.48\linewidth}
  \centering
  \centerline{\includegraphics[width=4.0cm]{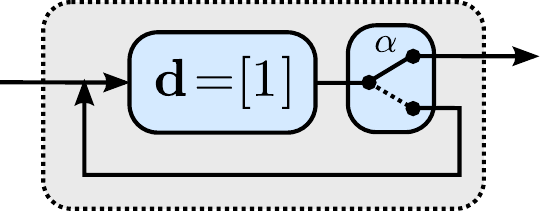}}
  \centerline{(d) Exponential mix as IIR.}\medskip
  \centerline{\includegraphics[width=4.0cm]{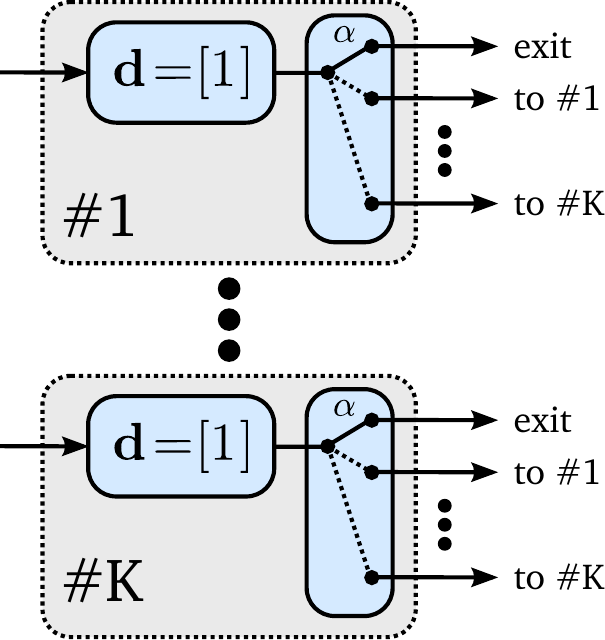}}
  \centerline{(e) Distributed implementation.}\medskip
\end{minipage}
\caption{Use of filter properties in anonymous network design.}
\label{fig:examples}
\end{figure}

 \textbf{a) Timed mix as a cascade of nodes.} We want to implement a time mix in a distributed way using 5 nodes in cascade, to protect $N=100$ senders against a short-term adversary that observes the inputs and outputs for $\rho=500$ rounds. The delay introduced by two nodes in cascade properly synchronized with the timer is the convolution of the individual delay characteristics (as in filters in cascade). Therefore, configuring the delay of the nodes as shown in Fig.~\ref{fig:examples}.a, we achieve the overall delay characteristic, result of convolving the 5 individual responses, shown in Fig.~\ref{fig:examples}.b (achieved). We see that this overall delay is close to the optimal one (objective). Figure~\ref{fig:examples}.c shows the frequency response, where we see that the nodes try to reduce the $\rho-N=400$ high-frequency DFT coefficients.
 
 \textbf{b) Distributed exponential mix.} We want to build an exponential mix, i.e., a timed mix with $\fpool{k}=\alpha(1-\alpha)^k$, which is optimal for some privacy metrics~\cite{DanContMixes,RebolloPool}. This is equivalent to a first-order IIR filter which we can implement easily in direct form, using a node that only delays messages until the end of a round ($\fpoolv=1$) and a switch that sends messages to an output with probability $\alpha$, and to the other with probability $1-\alpha$ (Fig.~\ref{fig:examples}.d). In order to implement this device in a decentralized way, we place several IIR filters in parallel as shown in Fig.~\ref{fig:examples}.e, where a message is forwarded to its recipient with probability $\alpha$ in each round, and fed to a random node with probability $1-\alpha$. The overall response is the same since this is equivalent to an scenario where we have identical IIR filters in parallel.
 
 \section{Conclusions}
 \label{sec:conclusions}
 
 In this work, we have studied the problem of designing the delay characteristic of timed mixes against a global adversary that aims at learning the communication patterns of the users. We have obtained theoretical solutions for this problem against a long-term and short-term attack, and have shown through experiments that they are close to the ones obtained numerically. We also showed the connection between designing delay characteristics and filters, and used this connection to implement the timed mixes in a decentralized way as a network of delaying nodes.

 \section{Appendix}
We provide an expression for the overall MSE $\MSEt$ of the best linear estimator of $\P$, described in \cite{ICASSP}, under the following conditions:
\begin{enumerate}
 \item The number of rounds observed by the adversary goes to infinity ($\rho\to\infty$) and it is much larger than the number of users in the system ($\rho\gg N$).
 \item The input processes are i.i.d.~as a Poisson distribution, i.e., $X_i^r\sim P(\srate{i})$.
 \item The average number of messages sent each round by all the users is much larger than one, i.e., $\sum_{i=1}^N\srate{i}\gg 1$.
\end{enumerate}
The expression we obtain only depends on the delay characteristic $\fpoolv\doteq[\fpool{0},\fpool{1},\cdots,\fpool{\rho-1}]^T$ through the following parameters: 
\begin{align}
 \gamma_1&\doteq \sum_k \fpool{k}^2\\
 \gamma_2&\doteq \sum_r \left(\sum_k \fpool{r} \fpool{r+k} \right)^2\\
 \gamma_3&\doteq\sum_k\fpool{k}^3.
\end{align}

After obtaining an expression for $\MSEt$, we prove that the MSE grows with $1/\gamma_1$ when the ``sharpness'' of each sender, defined as $\sharpi{i}\doteq\sum_{j=1}^M p_{j,i}^2$ for sender $i$, is almost zero, i.e., $\sharpi{i}\approx 0$, for all $i\in\{1,\cdots,N\}$. We also prove that the overall MSE grows with $(\gamma_1-\gamma_2)/\gamma_1^2$ when $\sharpi{i}\approx 1$ for all $i$.

\section{Theoretical expression for $\MSEt$.}
From \cite{ICASSP}, we get that
\begin{equation} \label{eq:MSEt}
 \Fitcol{\MSEt=\Exp{\trace{\M(\X^T\D^T\D\X)^{-1}\X^T\D^T\mathbf{\Sigma}_{\N|\X}\D\X(\X^T\D^T\D\X)^{-1}\M}}\,,}
\end{equation}
where
\begin{equation}
 \mathbf{\Sigma}_\mathbf{\N|\X}=\diag{\D\X\mathbf{1}_N}-\D\cdot\diag{\X\sharp}\cdot\D^T\,.
\end{equation}

We define
\begin{equation}
 \Rxx\doteq\frac{1}{\rho}\X^T\D^T\D\X
\end{equation}
and
\begin{equation}
 Rxyx\doteq\frac{1}{\rho}\X^T\D^T\mathbf{\Sigma}_{\N|\X}\D\X\,,
\end{equation}
and note that \eqref{eq:MSEt} can be written as
\begin{equation}
 \MSEt=\Exp{\trace{\M\Rxx^{-1}\Rxyx\Rxx^{-1}\M}}\,.
\end{equation}
The entries of $\Rxx$ and $\Rxyx$ are sample averages over $\rho$, and therefore as $\rho$ grows they get closer to their expected value. Using that the the input samples in $\X$ are i.i.d.~Poissonian with rates $\srates\doteq[\srate{1},\cdots,\srate{N}]^T$, we can compute
\begin{equation} \label{eq:Rxx}
 \Rxx=\srates\srates^T+\gamma_1\cdot\diag{\srates}\,.
\end{equation}

On the other hand, we can expand $\Rxyx$ as
\begin{equation}
 \Fitcol{\Rxyx=\frac{1}{\rho}\X^T\D^T\diag{\D\X\mathbf{1}_N}\D\X-\frac{1}{\rho}\X^T\D^T\D\diag{\X\sharp}\D^T\D\X\,.}
\end{equation}
Let $\Rxyx'$ and $\Rxyx''$ be the first and second summands of this expression, respectively. These summands can be written, when $\rho\to\infty$, as
\begin{equation}
 \Rxyx'=\srates\srates^T\left(2\gamma_1+\sum_{i=1}^N\srate{i}\right)+\diag{\srates}\left(\gamma_3+\gamma_1\cdot\sum_{i=1}^N\srate{i}\right)\,,
\end{equation}
and
\begin{equation}
 \begin{array}{lcl}
 \Rxyx''&=&\displaystyle\srates\srates^T\cdot\sum_{i=1}^N\srate{i}\sharpi{i}+\gamma_1\cdot\left[(\srates\circ\sharp)\srates^T+\srates(\srates\circ\sharp)^T\right]\\
 &+&\displaystyle\gamma_2\cdot\diag{\srates}\cdot\sum_{i=1}^N \srate{i}\sharpi{i}+\gamma_1^2\cdot\diag{\srates\circ\sharp}\,.
 \end{array}
\end{equation}
where $\circ$ is the entry-wise or Hadamard product.

In order to compute $\MSEt$, we need an expression for $\Rxx^{-1}$. Using the Sherman-Morrison formula in \eqref{eq:Rxx}, we can write
\begin{equation}
 \Rxx^{-1}=\frac{1}{\gamma_1}\left(\diag{\srates}^{-1}-\frac{\mathbf{1}_N\mathbf{1}_N^T}{\gamma_1+\sum_{i=1}^N\srate{i}}\right)\,.
\end{equation}
We then use our assumption $\sum_{i=1}^N\srate{i}\gg 1$ and the fact that $1\geq \gamma_1$ to approximate $\gamma_1+\sum_{i=1}^N\srate{i}\approx\sum_{i=1}^N\srate{i}$ in this expression.

Finally, we perform the matrix multiplications to obtain $\M\Rxx^{-1}\Rxyx\Rxx^{-1}\M$ and compute its trace to obtain a closed-form expression for $\MSEt$:
\begin{equation} \label{eq:general}
\begin{array}{lcl} 
 \MSEt&\approx&\displaystyle\frac{1}{\rho} \cdot \frac{1}{\gamma_1^2} \cdot \left(\gamma_1\cdot\sum_{i=1}^N\srate{i}-\gamma_2\cdot\sum_{i=1}^N\srate{i}\sharpi{i}+\gamma_3\right)\cdot \vspace{0.5cm}\\
 &&\displaystyle\left[\sum_{i=1}^N\srate{i}-\frac{\sum_{i=1}^N\srate{i}^2}{\sum_{i=1}^N\srate{i}}\right]\vspace{0.5cm}\\
 &+&\displaystyle\frac{1}{\rho}\cdot\left[\left(\frac{\sum_{i=1}^N\srate{i}^2}{(\sum_{i=1}^N\srate{i})^2}+1\right)\cdot\sum_{i=1}^N\srate{i}\sharpi{i}-\frac{\sum_{i=1}^N\srate{i}^2\sharpi{i}}{\sum_{i=1}^N\srate{i}}\right]\,.
\end{array}
\end{equation}
We study now the dependence of $\MSEt$ on the delay characteristic when $\sharpi{i}\approx 0$ and $\sharpi{i}\approx 1$. Note that, regardless of the value of $\sharpi{i}$, the second term in \eqref{eq:general} does not depend on the delay characteristic, so we can disregard it when studying how to design the delay characteristic to increase the MSE.

\section{Dependence of $\MSEt$ on the delay characteristic}

\subsection{First scenario ($\sharpi{i}\approx 0$).}
In this case, we can write
\begin{align}
 \gamma_1&\cdot\sum_{i=1}^N\srate{i}-\gamma_2\cdot\sum_{i=1}^N\srate{i}\sharpi{i}+\gamma_3\\
 &\approx\gamma_1\cdot\sum_{i=1}^N\srate{i}+\gamma_3\approx\gamma_1\cdot\sum_{i=1}^N\srate{i}\,,
\end{align}
where the first step comes from $\sharpi{i}\approx 0$ and the second one from $\gamma_3\leq\gamma_1$ and $\sum_{i=1}^N\srate{i}\gg 1$. Since the second term of \eqref{eq:general} can be disregarded when $\sharpi{i}\approx 0$, we have
\begin{equation}
 \MSEt\approx\frac{1}{\rho}\cdot\frac{1}{\gamma_1}\cdot \sum_{i=1}^N\srate{i} \cdot\left[\sum_{i=1}^N\srate{i}-\frac{\sum_{i=1}^N\srate{i}^2}{\sum_{i=1}^N\srate{i}}\right]\,.
\end{equation}
Then, the overall MSE of the adversary is proportional to $1/\gamma_1$, and therefore in order to increase $\MSEt$ we must increase $1/\gamma_1$.

\subsection{Second scenario ($\sharpi{i}\approx 1$).}
Here, by evaluating $\sharpi{i}\approx1$ and using the same approximations above, we get
\begin{equation}
 \MSEt\approx\frac{1}{\rho}\cdot \sum_{i=1}^N\srate{i} \cdot\left[\frac{\gamma_1-\gamma_2}{\gamma_1^2}\cdot\left(\sum_{i=1}^N\srate{i}-\frac{\sum_{i=1}^N\srate{i}^2}{\sum_{i=1}^N\srate{i}}\right)+1\right]\,.
\end{equation}
We can see that, in order to increase $\MSEt$, we must increase $(\gamma_1-\gamma_2)/\gamma_1^2$.

This concludes the proof.
\vfill\pagebreak

\bibliographystyle{IEEEbib}
\bibliography{references}

\end{document}